\documentclass[10pt,conference]{IEEEtran}

\usepackage{epsfig}
\usepackage{graphicx}
\usepackage{cite}
\usepackage{amsmath}
\usepackage{amssymb}
\usepackage{caption}
\newcommand{\BEAS}{\begin{eqnarray*}}
\newcommand{\EEAS}{\end{eqnarray*}}
\newcommand{\BEA}{\begin{eqnarray}}
\newcommand{\EEA}{\end{eqnarray}}
\newcommand{\BEQ}{\begin{equation}}
\newcommand{\EEQ}{\end{equation}}
\newcommand{\BIT}{\begin{itemize}}
\newcommand{\EIT}{\end{itemize}}
\newcommand{\BNUM}{\begin{enumerate}}
\newcommand{\ENUM}{\end{enumerate}}

\newcommand{\BA}{\begin{array}}
\newcommand{\EA}{\end{array}}

\newcommand{\ie}{{\it i.e.}}

\newcommand{\ones}{\mathbf 1}

\newcommand{\reals}{{\mbox{\bf R}}}

\newcommand{\diag}{\mathop{\bf diag}}

\renewcommand{\QED}{~~\rule[-1pt]{8pt}{8pt}}

\newcommand{\dom}{\mathop{\bf dom}}

\newcommand{\argmax}{\mathop{\rm argmax}}

{\begin{quote}}{\end{quote}}

\makeatletter
\long\def\@makecaption#1#2{
   \vskip 9pt
   \begin{small}
   \setbox\@tempboxa\hbox{{\bf #1:} #2}
   \ifdim \wd\@tempboxa > 5.5in
        \begin{center}
        \begin{minipage}[t]{5.5in}
        \addtolength{\baselineskip}{-0.95pt}
        {\bf #1:} #2 \par
        \addtolength{\baselineskip}{0.95pt}
        \end{minipage}
        \end{center}
   \else
	\hbox to\hsize{\hfil\box\@tempboxa\hfil}
   \fi
   \end{small}\par
}
\makeatother

\newcounter{oursection}

\newcounter{lecture}

\newcommand{\comment}[1]  {}
\def\BE{\begin{equation}}
\def\EE{\end{equation}}
\def\BEA{\begin{eqnarray}}
\def\EEA{\end{eqnarray}}

\newcommand\etal{{\textsl{et al.\,\,}}}

\begin{document}

\title{Distributed Large Scale\\ Network Utility Maximization}

\author{
\authorblockN{Danny Bickson, Yoav Tock}
\authorblockA{IBM Haifa Research Lab\\
Mount Carmel, Haifa 31905, Israel\\
Email: \{dannybi,tock\}@il.ibm.com}
\and
\authorblockN{Argyris Zymnis, Stephen P. Boyd}
\authorblockA{Department of Electrical Engineering\\
Stanford University\\
350 Serra Mall, Packard 243,\\
Stanford CA 94305, USA \\
Email: \{azymnis,boyd\}@stanford.edu
\and
\authorblockN{Danny Dolev}
\authorblockA{School of Computer Science and Engineering\\
Hebrew University of Jerusalem\\
Jerusalem 91904, Israel\\
Email: dolev@cs.huji.ac.il}
}
}
\maketitle

\begin{abstract}
Recent work by Zymnis \etal proposes an efficient primal-dual interior-point method, using
a truncated Newton method, for solving the network utility maximization (NUM) problem. This method has shown superior performance relative to the traditional dual-decomposition approach.
Other recent work by Bickson \etal shows how to compute efficiently and distributively
the Newton step, which is the main computational bottleneck of the Newton method, utilizing the Gaussian belief propagation algorithm.

In the current work, we combine both approaches to create an efficient distributed algorithm for solving
the NUM problem. Unlike the work of Zymnis, which uses a centralized approach, our new algorithm is
easily distributed. Using an empirical evaluation we show that our new method outperforms previous
approaches, including the truncated Newton method and dual-decomposition methods. As an additional contribution,
this is the first work that evaluates the performance of the Gaussian belief propagation algorithm vs. the preconditioned conjugate gradient method, for a large scale problem.
\end{abstract}


\section{Introduction}
We consider a network that supports a set of flows,
each of which has a nonnegative flow rate, and an associated
utility function.
Each flow passes over a route, which is a subset of the edges of
the network. Each edge has a given capacity, which is the maximum
total traffic (the sum of the flow rates through it) it can support.
The network utility maximization (NUM) problem is to choose the
flow rates to maximize the total utility, while respecting the
edge capacity constraints \cite{Sri:04,Ber:98}.
We consider the case where all utility functions are concave, in which
case the NUM problem is a convex optimization problem.

A standard technique for solving NUM problems is based on
dual decomposition \cite{DaW:60,Shor:85}.
This approach yields fully decentralized algorithms, that can scale
to very large networks.
Dual decomposition was first applied to the NUM problem
in \cite{Kelly:97},
and has led to an extensive body of research on distributed algorithms
for network optimization \cite{Low:99,CLCD:07,PalChiang:06}
and new ways to interpret existing network protocols \cite{Low:03}.

Recent work by Zymnis \etal presented a specialized primal-dual interior-point
method for the NUM problem~\cite{NUM}. Each Newton step is computed using the
preconditioned conjugate gradient method (PCG). This proposed method had a significant
performance improvement over the dual decomposition approach, especially when the
network is congested. Furthermore, the method can handle utility functions which
are not strictly concave. The main drawback of the primal-dual method is that it is centralized,
while the dual decomposition methods are easily distributed.

Other recent work by Bickson \etal \cite{Allerton08-1} proposes an efficient way for computing the
Newton step, which is the main computational effort of the primal-dual interior-point method
using the Gaussian belief propagation (GaBP) algorithm, which is an efficient distributed
algorithm.

In the current paper we propose to combine both previous approaches. We present an efficient
primal-dual interior point method, where the Newton step computed in each iteration is computed
using the GaBP algorithm. Using extensive simulations with very large
scale networks we compare the performance of our novel method to previous approaches
including an interior-point method using PCG, and dual decomposition methods.
Despite of being distributed, our new construction exhibits significant performance improvements
over previous approaches.

Furthermore, we provide the first comparison of performance of
the GaBP algorithm vs. the PCG method. The PCG method is a state-of-the-art method used extensively in
large-scale optimization applications. Examples include $\ell_1$-regularized
logistic regression \cite{KKB:07}, gate sizing \cite{joshi:eml}, and slack allocation
\cite{joshi:eml2}.
Empirically, the GaBP algorithm is immune to numerical
problems with typically occur in the PCG method, while demonstrating a faster convergence. The only previous work comparing the performance of GaBP vs. PCG we are aware of is \cite{Sthesis}, which used a small example of $25$ nodes,
and the work of \cite{BibDB:Weiss01Correctness} which used a grid of $25 \times 25$ nodes.

We believe that our approach is general and not limited to the NUM problem. It could potentially
be used for the solution of other large scale distributed optimization problems.

This paper is organized as follows. Section \ref{s-probform} briefly overviews the NUM problem formulation.
Section \ref{s-prev} outlines previous algorithms for solving the NUM problem, including dual descent and
truncated Newton method. We present our new construction which utilizes the GaBP algorithm in Section \ref{s-new-const}. Section \ref{s-exp-res} provides simulation results comparing the performance of the GaBP based algorithm with
the previous approaches. We conclude in Section \ref{s-conc}.

\section{Problem formulation}
\label{s-probform}

There are $n$ flows in a network, each of which is associated with
a fixed route, \ie, some subset of $m$ links.
Each flow has a nonnegative
\emph{rate}, which we denote $f_1, \ldots, f_n$.
With the flow $j$ we associate a utility function $U_j:\reals \rightarrow
\reals$, which is concave and twice differentiable, with $\dom U_j
\subseteq \reals_+$.
The utility derived by a flow rate $f_j$ is given by $U_j(f_j)$.
The total utility associated with all the flows is then
$U(f)=U_1(f_1)+\cdots +U_n(f_n)$.

The total traffic on a link in the network
is the sum of the rates of all flows that utilize that link.
We can express the link traffic compactly using the
\emph{routing} or \emph{link-route} matrix $R \in\reals^{m\times n}$,
defined as
\[
R_{ij} = \left\{\begin{array}{ll}
1 & \mbox{flow $j$'s route passes over link $i$}\\
0 & \mbox{otherwise}.
\end{array}\right.
\]
Each link in the network has a (positive)
\emph{capacity} $c_1, \ldots, c_m$.
The traffic on a link cannot exceed its capacity, \ie, we have
$Rf \leq c$, where $\leq$ is used for componentwise inequality.

The NUM problem is to choose the
rates to maximize total utility, subject to the link capacity
and the nonnegativity constraints:
\BEQ\label{e-rate_control}
\begin{array}{ll}
\mbox{maximize} & U(f)\\
\mbox{subject to} & Rf \leq c, \quad f \geq 0,
\end{array}
\EEQ
with variable $f \in \reals^n$.
This is a convex optimization problem and can be solved by a
variety of methods. We say that
$f$ is {\em primal feasible} if it satisfies $Rf\leq c$, $f\geq 0$.

The dual of problem (\ref{e-rate_control}) is
\BEQ\label{e-rate_control_dual}
\begin{array}{ll}
\mbox{minimize} & \lambda^Tc + \sum_{j=1}^n (-U_j)^*(-r_j^T\lambda)\\
\mbox{subject to} & \lambda \geq 0,
\end{array}
\EEQ
where $\lambda \in \reals^m_+$ is the dual variable associated with
the capacity constraint of problem (\ref{e-rate_control}),
$r_j$ is the $j$th column of $R$ and $(-U_j)^*$ is the
conjugate of the negative $j$th utility function \cite[\S 3.3]{BoV:04},
\[
(-U_j)^*(a) = \sup_{x \geq 0} (ax+U_j(x)).
\]
We say that $\lambda$ is {\em dual feasible} if it satisfies
$\lambda \geq 0$ and $\lambda \in \cap_{j=1}^n \dom (-U_j)^*$.

\section{Previous work}
\label{s-prev}
In this section we give a brief overview of the dual-decomposition
method and the primal-dual interior point method proposed in~\cite{NUM}.
%

%

\subsection{Dual decomposition}

Dual decomposition \cite{DaW:60,Shor:85,Kelly:97,Low:99}
is a projected (sub)gradient algorithm
for solving problem (\ref{e-rate_control_dual}), in the case
when all utility functions are strictly concave.
We start with any positive $\lambda$, and repeatedly
carry out the update
\begin{eqnarray*}
f_j &:=& \argmax_{x\geq 0}
\left(U_j(x)- x (r_j^T\lambda) \right), \quad j=1, \ldots, n,\\
\lambda &:=& \left(\lambda-\alpha\left(c-Rf\right)\right)_+,
\end{eqnarray*}
where $\alpha>0$ is the step size, and $x_+$ denotes the entrywise
nonnegative part of the vector $x$.
It can be shown
that for small enough $\alpha$, $f$ and $\lambda$ will converge
to $f^\star$ and $\lambda^\star$, respectively, provided all $U_j$ are
differentiable and strictly concave.
The term $s=c-Rf$ appearing in the update is the \emph{slack} in
the link capacity constraints (and can have negative entries during
the algorithm execution).
It can be shown that the slack is exactly the gradient
of the dual objective function.

Dual decomposition is a distributed algorithm. Each flow is
updated based on information obtained from the links it
passes over, and each link dual variable is updated based
only on the flows that pass over it.


\subsection{Primal-dual interior point method}
The primal-dual interior-point method is based on using a Newton step,
applied to a suitably modified form of the optimality conditions.
The modification is parametrized by a parameter $t$, which is adjusted
during the algorithm based on progress, as measured by the actual
duality gap (if it is available) or a surrogate duality gap (when the
actual duality gap is not available).

We first describe the search direction.
We modify the complementary slackness conditions to obtain
the modified optimality conditions
\begin{eqnarray*}
-\nabla U(f) + R^T \lambda-\mu &=& 0 \\
\diag(\lambda)s &=& (1/t)\ones \\
\diag(\mu)f &=& (1/t)\ones,
\end{eqnarray*}
where $t>0$ is a parameter that sets the accuracy of the approximation.
(As $t \rightarrow \infty$, we recover the optimality conditions for
the NUM problem.)
Here we implicitly assume that $f,s,\lambda,\mu > 0$.
The modified optimality conditions can be compactly written
as $r_t(f,\lambda,\mu) = 0$, where
\[
r_t(f,\lambda,\mu) = \left[\begin{array}{c}
-\nabla U(f) + R^T \lambda - \mu \\
\diag(\lambda)s - (1/t)\ones\\
\diag(\mu)f -(1/t)\ones
\end{array}\right].
\]

The primal-dual search direction is the
Newton step for solving the nonlinear equations
$r_t(f,\lambda,\mu) = 0$. If $y = (f,\lambda,\mu)$
denotes the current point, the Newton step
$\Delta y = (\Delta f,\Delta \lambda,\Delta \mu)$
is characterized by the linear equations
\[
r_t(y+\Delta y) \approx r_t(y)+ r'_t(y)\Delta y = 0,
\]
which, written out in more detail, are
\small
\begin{equation}\label{e-primdualstep}
\left[\begin{array}{ccc}
-\nabla^2U(f)     &   R^T  & -I   \\
-\diag(\lambda)R &  \diag(s) & 0  \\
\diag(\mu) & 0 & \diag(f)
\end{array}\right]
\left[\begin{array}{c}
\Delta f \\ \Delta \lambda \\ \Delta \mu
\end{array}\right]
= -r_t(f,\lambda,\mu).
\end{equation}
\normalsize

During the algorithm, the parameter $t$ is increased,
as the primal and dual variables approach optimality.
When we have easy access to a dual feasible point during the algorithm,
we can make use of the exact duality gap $\eta$ to set the value of $t$;
in other cases, we can use the surrogate duality gap $\hat \eta$.

The primal-dual interior point algorithm is given in \cite[\S 11.7]{BoV:04},
\cite{Wri:97}.

The most expensive part of computing the primal-dual
search direction is solving equation (\ref{e-primdualstep}).
For problems of modest size,
\ie, with $m$ and $n$ no more than $10^4$,
it can be solved using direct methods such as a
sparse Cholesky decomposition.

For larger problem instances ~\cite{NUM} proposes to solve (\ref{e-primdualstep})
\emph{approximately}, using a preconditioned conjugate gradient (PCG)
algorithm \cite[\S 6.6]{Dem:97}, \cite[chap. 2]{CTK:95}, \cite[chap. 5]{NoW:99}.
When an iterative method is used to approximately solve
a Newton system, the algorithm
is referred to as an {\em inexact}, \emph{iterative},
or {\em approximate} Newton method
(see \cite[chap. 6]{CTK:95} and its references).
When an iterative method is used inside a primal-dual
interior-point method, the overall algorithm is called a \emph{truncated-Newton primal-dual
interior-point method}. For details of the PCG algorithm,
we refer the reader to the references cited above.
Each iteration requires multiplication of the matrix by a vector,
and a few vector inner products.

\section{Our new construction}
\label{s-new-const}
%

Previous work of Zymnis \etal~\cite{NUM} shows that when applying the interior-point Newton method to the NUM problem,
each Newton step involves a solution of Eq. \ref{e-primdualstep}, 
where the solution $( \Delta f, \Delta \lambda, \Delta \mu)^T$ is the Newton search direction.

Recent results by Bickson~\etal
\cite{Allerton,ISIT1} utilizes the GaBP algorithm as an efficient distributed
 algorithm for solving a system of linear equations. For utilizing the GaBP
algorithm, we first normalize Eq. (\ref{e-primdualstep}) by $(1,-1/\lambda,-1/\mu)$
to get the following equivalent system of linear equations:
\small
\BE \left[
  \begin{array}{ccc}
    -\nabla^2 U(f) & R^T & -I \\
    R & -\diag(s/\lambda) & 0 \\
    -I & 0 & \diag(f/\mu) \\
  \end{array}
\right]
\left[
  \begin{array}{c}
    \Delta f \\
    \Delta \lambda \\
    \Delta \mu \\
  \end{array}
\right] = -\hat{rt}(f,\lambda,\mu). \label{newton_step_num_sym} \EE
\normalsize
where $\hat{rt}(f,\lambda,\mu) = ((-\nabla U(f) + R^T - \mu)^T, (-s + (\lambda/t))^T,
(-f + (\mu/t))^T)^T.$
Note that the new system of linear equations is symmetric, a condition
required by the GaBP algorithm.

The formulation~(\ref{newton_step_num_sym}) allows us to shift the
linear system of equations from an algebraic to a probabilistic domain.
Instead of solving a deterministic vector-matrix linear equation,
we now solve an inference problem in a graphical model describing
a certain Gaussian distribution function. Following \cite{ISIT2}
we define the joint covariance matrix \small \BE A \triangleq \left[
  \begin{array}{ccc}
    -\nabla^2 U(f) & R^T & -I \\
    R & -\diag(s/\lambda) & 0 \\
    -I & 0 & \diag(f/\mu) \\
  \end{array}
\right] \label{C_cov} \EE \normalsize and the shift vector $b
\triangleq ((-\nabla U(f) + R^T - \mu)^T, (-s + (\lambda/t))^T,
(-f + (\mu/t))^T)^T $. We further denote the search direction $x = (\Delta f^T, \Delta \lambda^T, \Delta \mu^T)^T$.

Given the covariance matrix $A$ and the shift vector $b$, one
can write explicitly the Gaussian density function \BE p(x) \sim \exp(-1/2x^TAx + b^Tx) \nonumber \EE
Now, we are interested in computing the MAP assignment:
\BE x^*  = \argmax_{x} \exp(-1/2x^TAx + b^Tx) \nonumber \EE
The corresponding graph of the covariance matrix $A$ is $\mathcal{G}$, with edge potentials (`compatibility functions') $\psi_{ij}$ and self-potentials
(`evidence') $\phi_{i}$. These graph potentials are determined
according to the following pairwise factorization of the Gaussian
distribution $p(x) \propto
\prod_{i=1}^{n}\phi_{i}(x_{i})\prod_{\{i,j\}}\psi_{ij}(x_{i},x_{j}),$
        resulting in $ \psi_{ij}(x_{i},x_{j})\triangleq \exp(-x_{i}A_{ij}x_{j}),$ and
        $ \phi_{i}(x_{i}) \triangleq
        \exp\big(b_{i}x_{i}-A_{ii}x_{i}^{2}/2\big).$
        The set of edges $\{i,j\}$ corresponds to the set of
        non-zero entries in $A$ (Eq. \ref{C_cov}). Hence, we would like to calculate the
marginal densities, which must also be Gaussian,
\BE p(x_{i}) \propto \mathcal{N}(\mu_{i}=\{A^{-1}b\}_{i},P_{i}^{-1}=\{A^{-1}\}_{ii}),
\nonumber \EE where $\mu_{i}$ and $P_{i}$ are the
marginal mean and inverse variance (a.k.a. precision),
respectively. Recall that, according to \cite{ISIT2}, the inferred
mean $\mu$ is identical to the desired solution of
(Eq. \ref{newton_step_num_sym}). The GaBP update rules are summarized in
Table~\ref{tab_summary}. We use the notation $\mathbb{N}(i)$ as the set of
node $i$ graph neighbors, excluding $i$.

\begin{table}[ht!]
\centerline{ \small
\begin{tabular}{|c|c|l|}
  \hline
  \textbf{\#} & \textbf{Stage} & \textbf{Operation}\\
  \hline
  1. & \emph{Initialize} & Compute $P_{ii}=A_{ii}$ and $\mu_{ii}=b_{i}/A_{ii}$.\\
  && Set $P_{ki}=0$ and $\mu_{ki}=0$, $\forall {k} \in \mathbb{N}(i)$.\\ \hline
  2. & \emph{Iterate} & Propagate $P_{ki}$ and $\mu_{ki}$, $\forall {k} \in \mathbb{N}(i)$\;.\\
  & & Compute $P_{i\backslash j}=P_{ii}+\sum_{{k} \in \mathbb{N}(i) \backslash j} P_{ki}$\\
  & & $\mu_{i\backslash j} = P_{i\backslash j}^{-1}(P_{ii}\mu_{ii}+\sum_{k \in \mathbb{N}(i) \backslash j}
P_{ki}\mu_{ki})$.\\
  & & Compute $P_{ij} = -A_{ij}P_{i\backslash j}^{-1}A_{ji}$ \\
  & &  $\mu_{ij} = -P_{ij}^{-1}A_{ij}\mu_{i\backslash j}$.\\\hline
  3. & \emph{Check} & If $P_{ij}$ and $\mu_{ij}$ did not converge,\\
  & &  return to \#2. Else, continue to \#4.\\\hline
  4. & \emph{Infer} & $P_{i}=P_{ii}+\sum_{{k} \in \mathbb{N}(i)}
P_{ki}$ \\
& &  $\mu_{i}=P_{i}^{-1}(P_{ii}\mu_{ii}+\sum_{k \in \mathbb{N}(i)} P_{ki}\mu_{ki})$.\\
  \hline
  5. & \emph{Output} & $x_{i}= \mu_{i} $ \\\hline
\end{tabular}}
\vspace{0.5cm}
\caption{Computing $x = A^{-1}b$ via GaBP \cite{ISIT1}.} \label{tab_summary}
\end{table}
\normalsize

It is known that if GaBP converges, it results in exact inference
~\cite{Weiss}. Determining the exact region of convergence remain open research problems. All that is known is a sufficient
(but not necessary) condition
stating that GaBP converges when the spectral radius satisfies
\mbox{$\rho(|I_{K}-A|)<1$}~\cite{WS,JMLR}. A stricter sufficient
condition~\cite{Weiss}, determines that the matrix $A$ must be
diagonally dominant (\ie, $|A_{ii}|>\sum_{j\neq i}|A_{ij}| ,
\forall i$) in order for GaBP to converge. Recently, a new technique for forcing convergence for any column-dependent matrices is proposed 
in \cite{ISIT09-1}. An upper bound on convergence speed is given in \cite{Allerton08-1}.

\section{Experimental results}
\label{s-exp-res}

\subsection{Small experiment}
In our first example we look at the performance of our method
on a small network. 
The utility functions are all logarithmic, \ie,
$U_j(f_j) = \log f_j$.
There are $n=10^3$ flows, and $m=2\cdot 10^3$ links.
The elements of $R$ are chosen randomly and independently,
so that the average route length is $10$ links.
The link capacities $c_i$ are chosen independently from a
uniform distribution on $[0.1,1]$. For this particular example, there are about $10^4$
nonzero elements in $R$ ($0.5\%$ density).

We compare three different algorithms for solving the NUM problem:
The dual-decomposition method, a truncated Newton method via PCG and a customized Newton method via the GaBP solver.
Out of the examined algorithms, the Newton method is centralized, while the dual-decomposition and GaBP solver are distributed algorithms. The source code of our Matlab simulation is available on \cite{MatlabGABP}.
\begin{figure}[ht!]
\centering{
  \includegraphics[width=250pt,clip]{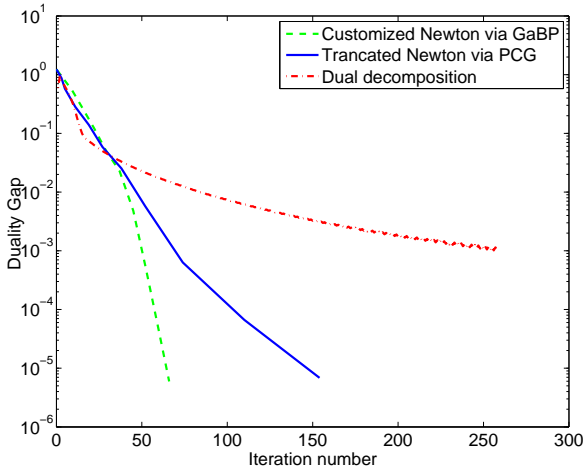}\\
  \caption{Convergence rate using the small settings.}\label{small_fig}
}
\end{figure}

Figure \ref{small_fig} depicts the solution quality, where the X-axis represents the number of algorithm
iterations, and the Y-axis is the surrogate duality gap (using a logarithmic scale).
As clearly shown, the GaBP algorithm has a comparable performance to the sparse Cholesky decomposition,
while it is a distributed algorithm. The dual decomposition method has much slower convergence.

\subsection{Larger experiment}
Our second example is too large to be solved using
the primal-dual interior-point method with direct search direction
computation, but is readily handled
by the truncated-Newton primal-dual algorithm using PCG,
the dual decomposition method and the customized Newton method via GaBP.
The utility functions are all logarithmic: $U_j(f_j) = \log f_j$.
There are $n=10^4$ flows, and $m=2\cdot 10^4$ links.
The elements of $R$ and $c$ are chosen as for the small example.
For dual decomposition, we initialized all $\lambda_i$ as $1$.
For the interior-point method, we initialized all $\lambda_i$ and $\mu_i$
as $1$. We initialize all $f_j$ as $\gamma$, where we choose $\gamma$
so that $Rf \leq 0.9 c$.

\begin{figure}[ht!]
  \includegraphics[width=250pt,clip]{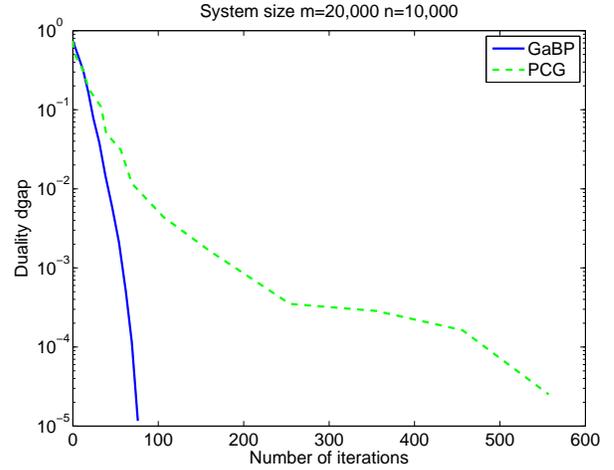}\\
  \caption{Convergence rate in the larger settings.}\label{larger_fig}
\end{figure}

Our experimental results shows, that
as the system size grows larger, the GaBP solver has favorable performance.
Figure \ref{larger_fig} plots the duality gap of both algorithms, vs. the number of iterations performed.

\begin{figure}[ht!]
  \includegraphics[width=250pt,clip]{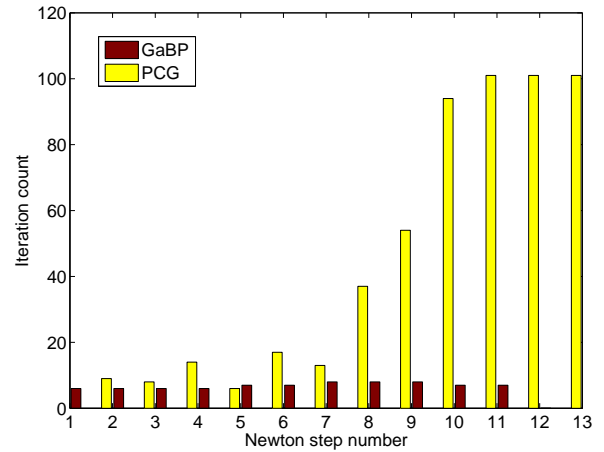}\\
  \caption{Iteration count per Newton step.}\label{gabp_vs_pcg}
\end{figure}

Figure \ref{gabp_vs_pcg} shows that in terms of Newton steps, both methods had comparable performance. The Newton method via the GaBP algorithm converged in 11 steps, to an accuracy of $10^{-4}$ where the truncated Newton method implemented via PCG converged in 13 steps to the same accuracy. However, when examining the iteration count in each Newton step (the Y-axis) we see that the GaBP remained constant, while the PCG iterations significantly increase as we are getting closer to the optimal point.

We have experimented with larger settings, up to $n=10^5$ flows, and $m=2\cdot 10^5$ links.
The GaBP algorithm converged in 11 Newton steps with 7-9 inner iteration in each Newton step.
The PCG method converged in 16 Newton steps with an average of 45 inner iterations.

\subsection{Numerical issues}
Overall, we have observed three types of numerical problems with
the PCG method. First, the PCG Matlab implementation runs into numerical problems and failed to compute
the search direction. Second, the line search failed, which means that no progress is possible in the computed direction without violating the problem constraints. Third, when getting close to the optimal solution, the number of PCG iterations significantly increases.

The numerical problems of the PCG algorithm are well known, see of example \cite{PCG1,PCG2}. In contrary, the GaBP algorithm did not suffer from the above numerical problems.

Furthermore, the PCG is harder to distribute, since in each PCG iteration a vector dot product and a matrix product
are performed. Those operations are global, unlike the GaBP which exploits the sparseness of the input matrix.

\section{Conclusion}
We propose an efficient distributed solution of the NUM problem using a customized Newton method, implemented via the GaBP algorithm. We compare the customized Newton method performance with state-of-the-art algorithms, including a dual descent method and a truncated Newton method, over large scale settings. We observe both faster convergence of the GaBP algorithm compared to both the preconditioned conjugate gradient and sparse Cholesky factorization. Furthermore, the GaBP does not suffer from numerical problems which affect the performance of the preconditioned conjugate gradient method.

We believe that the NUM problem serves as a case study for demonstrating the superior performance of the GaBP algorithm
in solving sparse systems of linear equations. Since the problem of solving a system of linear equations is a fundamental problem in computer science and engineering, we envision many other applications for our proposed method.

\label{s-conc}
\section*{Acknowledgment}
Danny Dolev is Incumbent of the Berthold Badler Chair in Computer Science. Danny Dolev was  supported in part by the Israeli Science Foundation (ISF) Grant number 0397373.

\bibliographystyle{IEEEtran}   
\bibliography{ISIT09-3,ls_num}       

\begin{thebibliography}{10}
\providecommand{\url}[1]{#1}
\csname url@rmstyle\endcsname
\providecommand{\newblock}{\relax}
\providecommand{\bibinfo}[2]{#2}
\providecommand\BIBentrySTDinterwordspacing{\spaceskip=0pt\relax}
\providecommand\BIBentryALTinterwordstretchfactor{4}
\providecommand\BIBentryALTinterwordspacing{\spaceskip=\fontdimen2\font plus
\BIBentryALTinterwordstretchfactor\fontdimen3\font minus
  \fontdimen4\font\relax}
\providecommand\BIBforeignlanguage[2]{{%
\expandafter\ifx\csname l@#1\endcsname\relax
\typeout{** WARNING: IEEEtran.bst: No hyphenation pattern has been}%
\typeout{** loaded for the language `#1'. Using the pattern for}%
\typeout{** the default language instead.}%
\else
\language=\csname l@#1\endcsname
\fi
#2}}

\bibitem{Sri:04}
R.~Srikant, \emph{The Mathematics of Internet Congestion Control}.\hskip 1em
  plus 0.5em minus 0.4em\relax Birkh\"{a}user, 2004.

\bibitem{Ber:98}
D.~Bertsekas, \emph{Network Optimization: Continuous and Discrete
  Models}.\hskip 1em plus 0.5em minus 0.4em\relax Athena Scientific, 1998.

\bibitem{DaW:60}
G.~B. Dantzig and P.~Wolfe, ``Decomposition principle for linear programs,''
  \emph{Operations Research}, vol.~8, pp. 101--111, 1960.

\bibitem{Shor:85}
N.~Z. Shor, \emph{Minimization Methods for Non-Differentiable Functions}.\hskip
  1em plus 0.5em minus 0.4em\relax Springer-Verlag, 1985.

\bibitem{Kelly:97}
F.~Kelly, A.~Maulloo, and D.~Tan, ``Rate control for communication networks:
  Shadow prices, proportional fairness and stability,'' \emph{Journal of the
  Operational Research Society}, vol.~49, pp. 237--252, 1997.

\bibitem{Low:99}
S.~H. Low and D.~E. Lapsley, ``Optimization flow control {I}: Basic algorithms
  and convergence,'' \emph{IEEE/ACM Transactions on Networking}, vol.~7, no.~6,
  pp. 861--874, Dec. 1999.

\bibitem{CLCD:07}
M.~Chiang, S.~H. Low, A.~R. Calderbank, and J.~C. Doyle, ``Layering as
  optimization decomposition: A mathematical theory of network architectures,''
  \emph{Proceedings of the IEEE}, vol.~95, no.~1, pp. 255--312, Jan. 2007.

\bibitem{PalChiang:06}
D.~Palomar and M.~Chiang, ``A tutorial on decomposition methods and distributed
  network resource allocation,'' \emph{IEEE Journal of Selected Areas in
  Communication}, vol.~24, no.~8, pp. 1439--1451, Aug. 2006.

\bibitem{Low:03}
S.~H. Low, ``A duality model of {TCP} and queue management algorithms,''
  \emph{IEEE/ACM Transactions on Networking}, vol.~11, no.~4, pp. 525--536,
  Aug. 2003.

\bibitem{NUM}
A.~Zymnis, N.~Trichakis, S.~Boyd, and D.~O'neill, ``An interior-point method
  for large scale network utility maximization,'' in \emph{Proceedings of the
  Allerton Conference on Communication, Control, and Computing}, 2007.

\bibitem{Allerton08-1}
D.~Bickson, Y.~Tock, D.~Dolev, and O.~Shental, ``Polynomial linear programming
  with {Gaussian} belief propagation,'' in \emph{the 46th {Allerton Conf. on
  Communications, Control and Computing}}, Monticello, IL, USA, 2008.

\bibitem{KKB:07}
K.~Koh, S.-J. Kim, and S.~Boyd, ``An interior point method for large-scale
  $\ell_1$-regularized logistic regression,'' \emph{Journal of Machine Learning
  Research}, vol.~8, pp. 1519--1555, July 2007.

\bibitem{joshi:eml}
S.~Joshi and S.~Boyd, ``An efficient method for large-scale gate sizing,''
  \emph{IEEE Trans. Circuits and Systems I: Fundamental Theory and
  Applications}, vol.~5, no.~9, pp. 2760--2773, Nov. 2008.

\bibitem{joshi:eml2}
------, ``An efficient method for large-scale slack allocation,'' in
  \emph{Engineering Optimization}, 2008.

\bibitem{Sthesis}
E.~B. Sudderth, ``Embedded trees: Estimation of {Gaussian} processes on graphs
  with cycles,'' Master's thesis, University of California at San Diego,
  February 2002.

\bibitem{BibDB:Weiss01Correctness}
Y.~Weiss and W.~T. Freeman, ``Correctness of belief propagation in {Gaussian}
  graphical models of arbitrary topology,'' \emph{Neural Computation}, vol.~13,
  no.~10, pp. 2173--2200, 2001.

\bibitem{BoV:04}
S.~Boyd and L.~Vandenberghe, \emph{Convex Optimization}.\hskip 1em plus 0.5em
  minus 0.4em\relax Cambridge University Press, 2004.

\bibitem{Wri:97}
S.~J. Wright, \emph{Primal-Dual Interior-Point Methods}.\hskip 1em plus 0.5em
  minus 0.4em\relax Society for Industrial and Applied Mathematics, 1997.

\bibitem{Dem:97}
J.~Demmel, \emph{Applied Numerical Linear Algebra}.\hskip 1em plus 0.5em minus
  0.4em\relax SIAM, 1997.

\bibitem{CTK:95}
C.~T. Kelley, \emph{Iterative Methods for Linear and Nonlinear
  Equations}.\hskip 1em plus 0.5em minus 0.4em\relax SIAM, 1995.

\bibitem{NoW:99}
J.~Nocedal and S.~J. Wright, \emph{Numerical Optimization}.\hskip 1em plus
  0.5em minus 0.4em\relax Springer, 1999.

\bibitem{Allerton}
D.~Bickson, O.~Shental, P.~H. Siegel, J.~K. Wolf, and D.~Dolev, ``Linear
  detection via belief propagation,'' in \emph{Proc. 45th Allerton Conf. on
  Communications, Control and Computing}, Monticello, IL, USA, Sept. 2007.

\bibitem{ISIT1}
O.~Shental, D.~Bickson, P.~H. Siegel, J.~K. Wolf, and D.~Dolev, ``Gaussian
  belief propagation solver for systems of linear equations,'' in \emph{IEEE
  Int. Symp. on Inform. Theory (ISIT)}, Toronto, Canada, July 2008.

\bibitem{ISIT2}
D.~Bickson, O.~Shental, P.~H. Siegel, J.~K. Wolf, and D.~Dolev, ``Gaussian
  belief propagation based multiuser detection,'' in \emph{IEEE Int. Symp. on
  Inform. Theory (ISIT)}, Toronto, Canada, July 2008.

\bibitem{Weiss}
Y.~Weiss and W.~T. Freeman, ``Correctness of belief propagation in {Gaussian}
  graphical models of arbitrary topology,'' \emph{Neural Computation}, vol.~13,
  no.~10, pp. 2173--2200, 2001.

\bibitem{WS}
J.~Johnson, D.~Malioutov, and A.~Willsky, ``Walk-sum interpretation and
  analysis of {Gaussian} belief propagation,'' in \emph{NIPS 05'}.

\bibitem{JMLR}
D.~M. Malioutov, J.~K. Johnson, and A.~S. Willsky, ``Walk-sums and belief
  propagation in {Gaussian} graphical models,'' in \emph{Journal of Machine
  Learning Research}, vol.~7, Oct. 2006.

\bibitem{ISIT09-1}
J.~K. Johnson, D.~Bickson, and D.~Dolev, ``Fixing convergence of {G}aussian
  belief propagation,'' in \emph{International Symposium on Information Theory
  (ISIT)}, Seoul, South Korea, July 2009.

\bibitem{MatlabGABP}
{Gaussian Belief Propagation} implementation in matlab [online] {\tt
  http://www.cs.huji.ac.il/labs/danss/p2p/gabp/}.

\bibitem{PCG1}
K.~C. Toh, ``Solving large scale semidefinite programs via an iterative solver
  on the augmented systems,'' \emph{SIAM J. on Optimization}, vol.~14, no.~3,
  pp. 670--698, 2003.

\bibitem{PCG2}
M.~Ko\v{c}vara and M.~Stingl, ``On the solution of large-scale {SDP} problems
  by the modified barrier method using iterative solvers,'' \emph{Math.
  Program.}, vol. 109, no.~2, pp. 413--444, 2007.

\end{thebibliography}

\end{document}